\documentclass[pra,aps,superscriptaddress,twocolumn,floatfix,nofootinbib,groupedaddress]{revtex4-2}

\usepackage{mathtools}
\usepackage[usenames,dvipsnames]{xcolor}
\usepackage[normalem]{ulem}
\usepackage{amsmath}
\usepackage{amssymb,mathrsfs}
\usepackage{graphicx}
\usepackage[colorlinks=true]{hyperref}
\usepackage{braket}

\begin{document}

\title{Qudit encoding in Rydberg blockaded arrays of atoms}

\author{Achille Robert}
\affiliation{Centre Européen de Sciences Quantiques and Institut de Science et d’Ingénierie Supramoléculaires (UMR 7006), University of Strasbourg and Centre National de la Recherche Scientifique, Strasbourg, France}

\author{Tom Bienaimé}
\email{t.bienaime@unistra.fr}
\affiliation{Centre Européen de Sciences Quantiques and Institut de Science et d’Ingénierie Supramoléculaires (UMR 7006), University of Strasbourg and Centre National de la Recherche Scientifique, Strasbourg, France}

\begin{abstract}
We propose a protocol to realize arbitrary state synthesis and unitary operations on a qudit encoded in the collective dressed states of a Rydberg-blockaded array of three-level atoms. This system is isomorphic to the Jaynes-Cummings model and acts as a multilevel Rydberg superatom whose nonlinear spectrum can be precisely controlled through the parameters of the laser driving the intermediate-to-Rydberg transition. Control of the qudit state is possible through pulse sequences of the laser driving the ground-to-intermediate transition. The dimension of the qudit Hilbert space is scalable by adjusting the number of atoms involved in the Rydberg-blockaded array. We estimate the fidelity of our protocol for realizing arbitrary unitaries and discuss the influence of the finite lifetime of the Rydberg state. Our work paves the way for processing quantum information with Rydberg-blockaded arrays of atoms as an alternative to atom qubit arrays. 
\end{abstract}

\maketitle

\section{Introduction}

While the commonly followed approach for digital quantum computing relies on storing quantum information as quantum bits, alternative approaches based on multidimensional quantum memories -- qudits -- hold great promise to reach beyond the capabilities of qubit-based quantum technologies, including potentially more efficient quantum algorithms, enhanced information encoding capabilities, and improvements for quantum error-correcting codes \cite{Wang20}.  Recently controlling qudit states and realizing quantum logic gates have been realized on a variety of experimental platforms including in photonics \cite{Erhard20,Chi22}, trapped ions \cite{Ringbauer22,Hrmo23}, molecular spins \cite{Moreno18}, superconducting devices \cite{Cervera22}, semiconductor platforms \cite{Fernandez24}, and cold atoms \cite{Chaudhury07} or molecules \cite{Sawant20}.

In this study, we consider neutral atoms which are rapidly becoming a leading platform for digital quantum computing \cite{Henriet20,Morgado21}. As an alternative,  to single-atom qubit encoding, it has recently been demonstrated that a Rydberg-blockaded ensemble of cold atoms, known as a \emph{Rydberg superatom}, can encode a qubit in two collective states. Innovative experimental approaches have been followed to create a well-controlled Rydberg superatom including atoms from a thermal cloud loaded into an optical lattice \cite{Mei22}, a two-dimensional Mott insulator \cite{Zeiher15}, single atoms in optical tweezers \cite{Labuhn16}, and a system of defect-free optical tweezers \cite{Bernien17}. The Rydberg-superatom approach enables fast qubit readout \cite{Xu21} and several superatom-based quantum computing methods have been recently introduced \cite{Cesa23, Byun24}. Here, we investigate using a Rydberg-blockaded array of three-level atoms to encode quantum information in its collective dressed states enabling the use of this system as a qudit. This system extends the concept of a two-level Rydberg superatom to the one of a multilevel Rydberg superatom which can be spectroscopically controlled by exploiting the nonlinearity of its spectrum.

When a laser resonantly couples the intermediate and Rydberg states of individual three-level atoms arranged in a Rydberg-blockaded array, the resulting collective dressed states (polaritons) are closely related to those of the Jaynes-Cummings model \cite{Jaynes63,Greentree13}, as well as to the broader spin-boson model, as demonstrated theoretically \cite{Keating16} and experimentally \cite{Lee17}. While the standard Jaynes-Cummings model describes a single two-level system interacting with a single quantized bosonic mode, our Rydberg-superatom implementation realizes the same effective model using a different physical platform. More specifically, these dynamics are realized without a physical cavity, using laser-driven atomic transitions. The effective light-matter coupling strength is directly controlled via external laser fields, allowing flexible and potentially time-dependent tuning. The effective bosonic mode is associated with collective atomic excitations in the intermediate state, with a linewidth determined by the underlying atomic states rather than by a cavity decay rate. The spin degree of freedom corresponds to the presence or absence of a single Rydberg excitation in the blockaded ensemble, forming a collective spin-1/2-like system. These features provide a flexible and fully controllable atomic platform for implementing Jaynes-Cummings physics.

The whole Rydberg-blockaded atomic array can be viewed as a multilevel Rydberg superatom with controllable energy levels whose splittings can be tuned both in magnitude and sign by adjusting the parameters of the laser driving the upper atomic transition. This approach enables one to scale up the qudit dimension simply by adding more atoms to the system. Qudits encoded in the Hilbert space spanned by the dressed states are controlled by a pulse sequence of the laser or the radio-frequency field coupling the atomic ground and intermediate states.  This enables one to synthesize any target state and realize arbitrary unitary operations on the qudit Hilbert space. Exact pulse sequences can be computed \emph{ab initio} even for very complex operations and could be used as a good starting point for optimal control methods. Our protocol only requires global spectroscopic addressing of the whole array of Rydberg-blockaded individual atoms which is a significant advantage over techniques that require site-selective addressing of the atomic array. This work is complementary to recent studies that propose to take advantage of the Jaynes-Cummings model for applications in quantum information science \cite{Mischuck13,Keating16,Villas19} and  goes beyond a previous theoretical study demonstrating the full control over the Hilbert space using optimal control techniques \cite{Keating16} or an earlier proposition for realizing arbitrary unitaries with the Jaynes-Cummings model \cite{Mischuck13}.

\section{The model}

\begin{figure*}[t!]
\centering
\includegraphics[width=\linewidth]{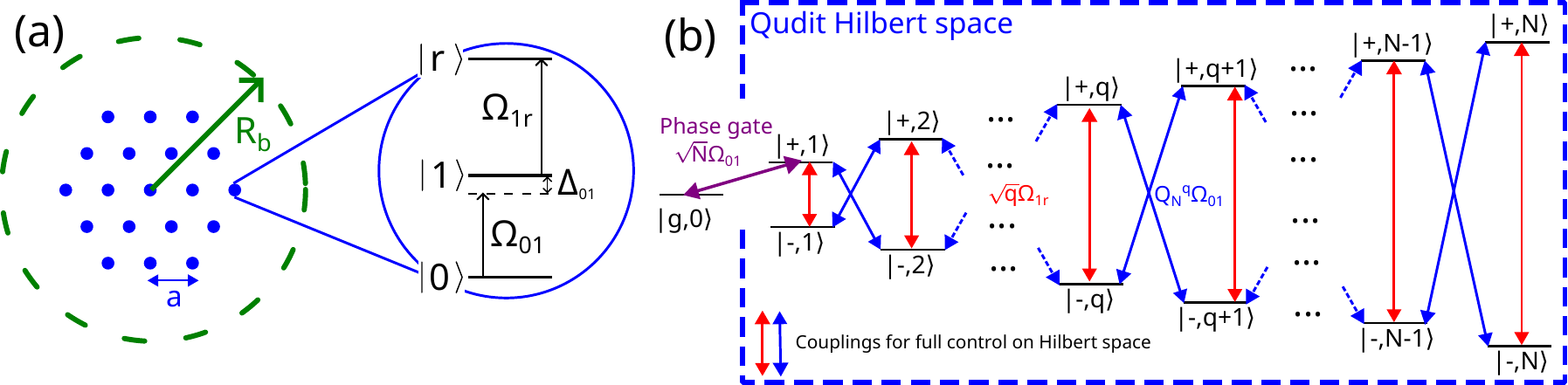}
	\caption{\textbf{Rydberg-blockaded array of single atoms and its collective dressed-state level scheme}. \textbf{(a)} $N$ individual three-level atoms are positioned regularly either using an array of optical tweezers or by loading the sites of an optical lattice with a tweezer. The atoms feature a ladder-type level structure, where the upper level corresponds to a Rydberg state. The interatomic distance $a$ is chosen such that radiative dipole-dipole interactions associated with the optical transitions can be neglected ($a > \lambda$, where $\lambda$ is the wavelength of the light driving either the upper or lower transition, if the latter lies within the optical domain), while ensuring that all atoms lie within a blockaded volume ($N^{1/d} a < R_b$, where $d$ is the dimensionality of the system). \textbf{(b)} Energy diagram of the dressed states, revealing the nonlinear Jaynes-Cummings ladder. Arrows represent the coupling terms of the Hamiltonian in the dressed-state basis.}\label{Fig1}
\end{figure*}

We consider an array of $N$ individual atoms located at positions $\mathbf x_j$. Each atom has three levels $\left\{|0_j\rangle,|1_j\rangle,|r_j\rangle \right\}$ in a ladder configuration where the level $|r_j\rangle$ is a Rydberg state subject to strong van der Waals interaction (see Fig. \ref{Fig1}).  Throughout this work, we set $\hbar = 1$ and express all energies in units of angular frequency. A laser resonantly couples $|1_j\rangle \leftrightarrow |r_j\rangle$, leading to the bare system Hamiltonian
\begin{align} \label{H_0}
	\hat H_{0} = &  \frac{\Omega_{1r}}{2} \sum_j \left[  e^{-i \phi_{1r}}   |r_j\rangle \langle 1_j| +  e^{i \phi_{1r}}     |1_j\rangle \langle r_j| \right] \notag \\                                                         
        & + \frac{1}{2}  \sum_j \sum_{k\neq j} V_{jk}  |r_j\rangle \langle r_j| \otimes |r_k\rangle \langle r_k| \, ,   
\end{align}
where $\Omega_{1r}$ is the Rabi frequency, $\phi_{1r}$ is the phase of the laser, and $V_{jk} = C_6/||\mathbf x_j - \mathbf x_k||^6$ is the interatomic interaction potential. We suppose that all atoms are located inside a blockaded sphere where $R_{\text{b}} = \left(|C_6|/\Omega_{1r}\right)^{1/6}$ is the blockade radius. This condition implies that at most one atom can be excited to a Rydberg state (hard-blockade regime). The Hamiltonian maps onto the Jaynes-Cummings model \cite{Keating16} and can then be conveniently written as (see appendix \ref{Appendix_A} for the detailed derivation)
\begin{align} \label{H_0_dressed_states}
        \hat H_{0} = &  \sum_{q=1}^{N} \, \frac{\Omega_{1r} \sqrt{q}}{2}    \, \mathbf n_{\phi_{1r}} \cdot \widehat{\boldsymbol{\sigma}}_{\left\{|+,q\rangle,|-,q\rangle \right\}} \, ,   
\end{align}
where the vectorial Pauli operators $\widehat{\boldsymbol{\sigma}}_{\left\{|u\rangle,|v\rangle \right\}} = \left[\hat \sigma_x, \hat \sigma_y, \hat \sigma_z\right]$ act on the subspace $\left\{|u\rangle,|v\rangle \right\}$, and the unit vector is $\mathbf n_{\phi_{1r}} = (0, - \sin \phi_{1r},\cos \phi_{1r})$. We have introduced the dressed states $|\pm,q\rangle = \frac{1}{\sqrt 2} \left( |e,q-1\rangle \pm |g,q\rangle \right)$ with $\ket{g,q} = \{\ket{0}^{\otimes N-q}\ket{1}^{\otimes q}\}_{\text{sym}}$ and $\ket{e,q} = \{\ket{0}^{\otimes N-q-1}\ket{1}^{\otimes q}\ket{r}\}_{\text{sym}}$. Note that for $\phi_{1r} = 0$ or $\pi$, the Hamiltonian is diagonal in the basis $\left\{|\pm,q\rangle\right\}$ with the well-known nonlinear Jaynes-Cummings ladder spectrum $E_{\pm,q}[\phi_{1r} = 0] = \pm  \frac{\Omega_{1r} \sqrt{q}}{2}$ and  $E_{\pm,q}[\phi_{1r} = \pi] = \mp  \frac{\Omega_{1r} \sqrt{q}}{2}$. This Rydberg-blockaded array of $N$ single atoms constitutes a fully controllable multilevel Rydberg superatom with $2N+1$ levels $\left\{|g,0\rangle,|\pm,q\rangle\right\}$ for which the energy splitting can be tuned both in magnitude and sign by adjusting the parameters of the laser of the upper transition $\left(\Omega_{1r},\phi_{1r}\right)$.

To exploit these dressed states for encoding and processing quantum information, we introduce a laser that couples $|0_j\rangle \leftrightarrow |1_j\rangle$, thereby enabling transitions via the control Hamiltonian:
\begin{align} \label{H_c}
	\hat H_{\text{c}} = &  \frac{\Omega_{01}}{2} \sum_j \left[  e^{-i \phi_{01}}   |1_j\rangle \langle 0_j| +  e^{i \phi_{01}}     |0_j\rangle \langle 1_j| \right] \notag \\
	& - \Delta_{01} \sum_j \left[ |1_j\rangle \langle 1_j|  + |r_j\rangle \langle r_j|\right]
\end{align}
where $\Omega_{01}$ is the Rabi frequency, $\phi_{01}$ the phase of the laser, and $\Delta_{01}$ the detuning of the laser with respect to the $|0\rangle \leftrightarrow |1\rangle$ transition. After some algebra, the control Hamiltonian can be written in the dressed-state basis (see appendix \ref{Appendix_A} for the detailed derivation) 
\begin{align} \label{H_c_dressed_states}
\hat{H}_{\text{c}} ={} & \frac{\Omega_{01}}{2} \Biggl[ \sum_{\pm} \sum_{q=1}^{N-1} K_{N}^{q} \, \mathbf{n}_{\phi_{01}} \cdot \widehat{\boldsymbol{\sigma}}_{\{|\pm,q+1\rangle,|\pm,q\rangle\}} \notag \\
& \quad - \sum_{\pm} \sum_{q=1}^{N-1} Q_{N}^{q} \, \mathbf{n}_{\phi_{01}} \cdot \widehat{\boldsymbol{\sigma}}_{\{|\pm,q+1\rangle,|\mp,q\rangle\}} \notag \\
& \quad + \sqrt{\frac{N}{2}} \, \mathbf{n}_{\phi_{01}} \cdot \widehat{\boldsymbol{\sigma}}_{\{|+1\rangle,|g,0\rangle\}} \notag \\
	& \quad - \sqrt{\frac{N}{2}} \, \mathbf{n}_{\phi_{01}} \cdot \widehat{\boldsymbol{\sigma}}_{\{|-1\rangle,|g,0\rangle\}} \Biggr] \notag \\
& - \Delta_{01} \sum_{\pm} \sum_{q=1}^N q \, |\pm,q\rangle \langle \pm,q| \, ,
\end{align}
where $K_N^q = \frac{\sqrt{(N-q)}}{2 \left(\sqrt{q+1} - \sqrt{q}\right) }$, $Q_N^q = \frac{\sqrt{(N-q)}}{2 \left(\sqrt{q+1} + \sqrt{q}\right)}$, and $\mathbf n_{\phi_{01}} = (\cos \phi_{01}, \sin \phi_{01},0)$. 

After rewriting both $\hat H_0$ and $\hat H_c$ from the bare to the dressed-state basis, we define the total Hamiltonian
\begin{equation} \label{Htot}
\hat H = \hat H_0 + \hat H_c
\end{equation}
which governs the full dynamics of the system. This Hamiltonian provides the framework we use in the next section to process quantum information encoded in the dressed states.

From a theoretical standpoint, the atoms could be randomly positioned as long as they all fit inside a blockaded sphere, as in an ensemble within a single optical tweezer. However, in that case, the uncertainty in the number of atoms makes precise control of the qudit state nearly impossible, since the Hamiltonian depends on collective parameters. Furthermore, atomic ensembles are likely affected by light-assisted collisional loss processes during qudit manipulation \cite{Schlosser02}, or by cooperative effects arising from dipole-dipole radiative couplings \cite{Bienaime13}. To overcome these limitations, we propose microstructuring the ensemble by positioning single atoms at regular intervals, either using an array of optical tweezers \cite{Endres16, Barredo16} or by loading specific sites of an optical lattice with optical tweezers, as illustrated in Fig. \ref{Fig1}. 

\section{Gate protocol}  \label{Gate_Protocol}

We encode quantum information in the Hilbert space spanned by the dressed-states basis $\left\{|\pm,q\rangle\right\}_{q=1..N}$ (dimension $2N$) such that any qudit state vector can be written as $|\psi\rangle =  \sum_q \sum_\pm a_{\pm,q} |\pm,q\rangle$. We denote by $\mathcal H$ the Hilbert space spanned by $\left\{|g,0\rangle,|\pm,q\rangle\right\}_{q=1..N}$ and by $\mathcal H'$ the qudit Hilbert space spanned by $\left\{|\pm,q\rangle\right\}_{q=1..N}$. In the following, we show how to process quantum information on such a platform by demonstrating the ability to (i) prepare any initial state, (ii) realize arbitrary unitary operations on the qudit Hilbert space, and (iii) characterize the final state by projective measurements. All these steps are achieved through a series of pulses $\left\{T^{(k)},\Omega_{1r}^{(k)},\phi_{1r}^{(k)},\Omega_{01}^{(k)},\phi_{01}^{(k)},\Delta_{01}^{(k)} \right\}$, where $T^{(k)}$ is the duration of the $k$th pulse.

\subsection{Arbitrary qudit gate}

We follow the method of Ref. \cite{Muthukrishnan00} to realize arbitrary quantum logic on the qudit Hilbert space. Any unitary operator acting on $\mathcal H'$ can be decomposed as $\hat U = \sum_{j=1}^{2N} e^{i \alpha_j} |\alpha_j\rangle \langle\alpha_j|$ where $e^{i \alpha_j}$  and $|\alpha_j\rangle$ are the eigenvalues and the eigenvectors of $\hat U$ respectively. It can then be obtained by subsequently applying the pulse sequence that creates a generalized phase gate with angle $\alpha_j$ on the target vector $|\alpha_j\rangle$, i.e., $\hat U = \prod_{j=1}^{2N} \hat P_{|\alpha_j\rangle, \alpha_j}$ where $\hat P_{|\psi\rangle, \Phi}$ is the generalized phase gate acting on the target vector $|\psi\rangle$ with angle $\Phi$. This generalized phase gate is realized using two key ingredients: (i) the full control over the Hilbert space $\mathcal H'$, denoted by $\hat O_{|\psi\rangle}$, that maps any $|\psi\rangle \in \mathcal H'$ to $|-,1\rangle$ and (ii) the generalized phase gate $\hat P_{|-,1\rangle, \Phi}$ which applies a phase gate with angle $\Phi$ to the state $|-,1\rangle$. Combining these two operations gives $\hat P_{|\psi\rangle, \Phi} =  \hat O_{|\psi\rangle}^{-1}   \hat P_{|-,1\rangle, \Phi}  \hat O_{|\psi\rangle}$ and can be used to realize any unitary operation.

\subsection{Full control over the Hilbert space}

The goal of this subsection is to describe how to realize the full control over the Hilbert space. First, we perform a series of rotations on the subspaces $\left\{|\pm,q+1\rangle, |\mp,q\rangle \right\}$ to fold the state vector $|\psi\rangle = \sum_\pm \sum_q a_{\pm,q} |\pm,q\rangle$ on the subspace spanned by $\left\{|+,1\rangle, |-,1\rangle\right\}$. To do this, we define the projector $\hat Q_{\left\{|\pm,q+1\rangle, |\mp,q\rangle\right\}}$ on the subspace $\left\{|\pm,q+1\rangle, |\mp,q\rangle\right\}$ to calculate the coordinates $\mathbf u = \langle \psi|\hat Q \widehat{\boldsymbol{\sigma}} \hat Q|\psi\rangle$ of the projected state $\hat Q |\psi\rangle$ on the Bloch sphere where the state $|\mp,q\rangle$ is pointing upward. To fold $\hat Q |\psi\rangle$ onto $|\mp,q\rangle$ we set $\Delta_{01} = \pm \frac{\Omega_{1r}}{2\left( \sqrt{q+1} - \sqrt{q} \right) }$ and $\phi_{1r} = 0$ to get the effective approximate Hamiltonian (see appendix \ref{Appendix_B} for the detailed derivation)
\begin{align}
    \hat{H}_{\text{eff}} ={} & - \frac{\Omega_{01} Q_N^q}{2} \, \mathbf{n}_{\phi_{01}} \cdot \widehat{\boldsymbol{\sigma}}_{\{|\pm,q+1\rangle,|\mp,q\rangle\}} \notag \\
    & + \sum_\pm \sum_{q=1}^N \left(E_{\pm,q}[\phi_{1r}] - \Delta_{01} q \right) \, |\pm,q\rangle \langle \pm,q| \, . \label{eq:Heff}
\end{align}
We set $\phi_{01}$ such that $\mathbf n_{\phi_{01}} = \mathbf u \times \mathbf u_z$ and perform a rotation around this axis by an angle $\Omega_{01} Q_N^q T = \arccos \left(\mathbf u \cdot \mathbf u_z\right)$ which sets the evolution time $T$ for this operation. This rotation $\widehat{\mathcal R}_{\left\{|\pm,q+1\rangle, |\mp,q\rangle\right\}} = e^{-i \hat H_{\text{eff}} T}$ leaves the system in the state $|\psi_{\text{after}}\rangle =    \left( \hat I -  \hat Q \right)  \left\{ \sum_\pm \sum_q e^{i(E_{\pm,q} - q \Delta_{01}) t} |\pm,q\rangle \langle\pm,q| \right\}  |\psi\rangle  +   a_{\mp,q} \sqrt{1 + \frac{\left|a_{\pm,q+1} \right|^2}{\left|a_{\mp,q} \right|^2  } }    |\mp,q\rangle$. Note that, if needed, it is possible to cancel the phase accumulated during the rotation by applying two pulses such that $\widetilde{\mathcal R} =  e^{-i \hat H_{\text{eff}}\left[\phi_{1r} = \pi, -\Delta_{01} \right] \frac{T}{2}} e^{-i \hat H_{\text{eff}}\left[\phi_{1r} = 0, \Delta_{01} \right] \frac{T}{2}}$. By iterating these rotations in the proper order, $\prod_{q=1}^{N-1}  \prod_{\pm}  \widehat{\mathcal R}_{\left\{|\pm,N+1-q\rangle,|\mp,N-q\rangle \right\}}$, we end up in the subspace spanned by $\left\{|+,1\rangle, |-,1\rangle\right\}$. For the final stage to get full control over $\mathcal H'$, we cannot use the state $|g,0\rangle$ which lies outside of $\mathcal H'$. We combine two rotations in the subspace $\left\{|+,1\rangle,|-,1\rangle \right\}$ by turning off the $|0\rangle \leftrightarrow |1\rangle$ laser ($\Omega_{01} = 0$, $\Delta_{01} = 0$) so that the full Hamiltonian reduces to Eq. (\ref{H_0_dressed_states}). After finding the coordinates $(\theta,\phi)$ of the state on the Bloch sphere where the state $|-,1\rangle$ is pointing upward, we perform a first rotation around $- \mathbf u_z$ ($\phi_{1r} = \pi$) for a duration $T = \phi/ \Omega_{1r}$ followed by a rotation around $- \mathbf u_y$ ($\phi_{1r} = \pi/2$) for a duration $T = \theta / \Omega_{1r}$. After this whole pulse sequence, we end up in the state $|-,1\rangle$ hence completing the full control over $\mathcal H'$. The inverse operation $\hat O^{-1}_{|\psi\rangle}$ is obtained by reversing the order of the pulse sequence after applying (i) for the rotations on the $\left\{|\pm,q+1\rangle, |\mp,q\rangle \right\}$ subspaces -- $\phi_{01} \rightarrow \phi_{01} + \pi$ to invert the rotation axes, and $\phi_{1r} = 0 \rightarrow \pi$, $\Delta_{01} \rightarrow -\Delta_{01}$ for canceling the phase accumulation terms (not needed if one uses the alternative rotations $\widetilde{\mathcal R}$) -- and (ii) on the $\left\{|+,1\rangle,|-,1\rangle \right\}$ subspace, $\phi_{1r} \rightarrow \phi_{1r}  + \pi$ to reverse the rotation axes.

\subsection{Phase gate on $|-,1\rangle$}

This gate is realized by performing a $z$ rotation in the $\left\{|-,1\rangle,|g,0\rangle \right\}$ subspace by an angle $\Phi$, such that the state $|-,1\rangle$ transforms to $e^{i \Phi}|-,1\rangle$. As $\mathbf n_{\phi_{01}}$ lies in the $xy$ plane, the $z$ rotation is obtained by two subsequent $\pi$ rotations. First, we set  $\phi_{1r} = 0$, $\Delta_{01} = - \Omega_{1r}/2$ to realize a $\pi$ rotation around $ \mathbf n_{\phi_{01}} = (\cos \Phi, \sin \Phi, 0)$ by setting $\phi_{01} = \Phi$ and letting the system evolve during $T = \sqrt 2 \pi /(\sqrt N \, \Omega_{01})$ with the following effective Hamiltonian (see appendix \ref{Appendix_B} for the detailed derivation): 
\begin{align} \label{Heff_Phase_Gate}
	\hat H_{\text{eff}} = & - \sqrt{\frac{N}{2}} \frac{\Omega_{01} }{2} \, \mathbf n_{\phi_{01}} \cdot \widehat{\boldsymbol{\sigma}}_{\left\{|-,1\rangle,|g,0\rangle \right\}} \notag \\
	& + \sum_\pm \sum_{q=1}^N \left(E_{\pm,q}[\phi_{1r}] - \Delta_{01} q \right) \, |\pm,q\rangle \langle\pm,q| \, .
\end{align}
For the second $\pi$ pulse, we set $\phi_{1r} = \pi$, $\Delta_{01} = \Omega_{1r}/2$ in order to cancel the phase terms accumulated during the first $\pi$ pulse and perform a rotation around $\mathbf u_x$ by setting $\phi_{01} = 0$ during $T = \sqrt 2 \pi /(\sqrt N \, \Omega_{01})$. The combination of these two pulses realizes a phase gate on the state $|-,1\rangle$ by an angle $\Phi$. The state $|g,0\rangle$ (populated during the process but not at the beginning or at the end) enables one to realize a phase gate on $|-,1\rangle$ from a $z$ rotation of an angle $\Phi$ in the subspace $\left\{|-,1\rangle,|g,0\rangle \right\}$ which is the reason why we excluded the state $|g,0\rangle$ from the qudit Hilbert space $\mathcal H'$.

\section{Initialization and measurements}

This section describes how to prepare any initial state and how to realize projective measurements on the qudit Hilbert space.

\subsection{State initialization}

The ability to prepare any initial state $|\psi_{\text{ini}}\rangle \in \mathcal H'$ from the ground state $|g,0\rangle$ results from the \emph{full control over the Hilbert space} $\mathcal H$. The full control over $\mathcal H$ which sends any $|\psi\rangle \in \mathcal H \rightarrow |g,0\rangle$ is obtained in a similar way to the full control over $\mathcal H'$ that we detailed above by substituting the final rotation in the $\left\{|+,1\rangle,|-,1\rangle \right\}$ subspace by a first rotation in the subspace $\left\{|+,1\rangle,|g,0\rangle \right\}$ followed by a second one in $\left\{|-,1\rangle,|g,0\rangle \right\}$. 

\subsection{State measurement}

The nonlinear spectrum of the dressed states can be exploited to measure the qudit state using spectroscopic techniques. For example, coupling the atomic state $\ket{0}$ to an auxiliary state $\ket{2}$ via a cycling transition driven by a laser resonant with the $\ket{0}\leftrightarrow\ket{2}$ transition realizes a projective measurement onto the collective state $\ket{g,0}$. Experimentally, this is performed by collecting the fluorescence emitted on this transition. For $\phi_{1r}=0$, the dressed states $\ket{\pm,q}$, with energies $E_\pm=\pm \frac{\Omega_{1r}\sqrt{q}}{2}$, are detuned from the laser and therefore do not fluoresce. This projective measurement onto $\ket{g,0}$ can be extended to an arbitrary target state $\ket{\psi_{\mathrm{target}}}$ in the qudit Hilbert space. Exploiting full control over the Hilbert space $\mathcal H$, one applies a pulse sequence implementing the operator $\hat O_{\ket{\psi_{\mathrm{target}}}}$, which maps $\ket{\psi_{\mathrm{target}}}$ onto $\ket{g,0}$, to the system in state $\ket{\psi}$. A subsequent spectroscopic measurement of $\ket{g,0}$ thus realizes a projective measurement of $\ket{\psi}$ onto $\ket{\psi_{\mathrm{target}}}$, since $\langle g,0 | \hat O_{\ket{\psi_{\mathrm{target}}}} \ket{\psi} = \langle \psi_{\mathrm{target}} | \psi \rangle$. Accordingly, the fluorescence signal observed after applying $\hat O_{\ket{\psi_{\mathrm{target}}}}$ and illuminating the system with a resonant laser on the $\ket{0}\leftrightarrow\ket{2}$ transition is proportional to $|\langle \psi_{\mathrm{target}} | \psi \rangle|^2$, thereby realizing the desired projective measurement.

\section{Applications}

\begin{figure}[t!]
\centering
	\includegraphics[width=\linewidth]{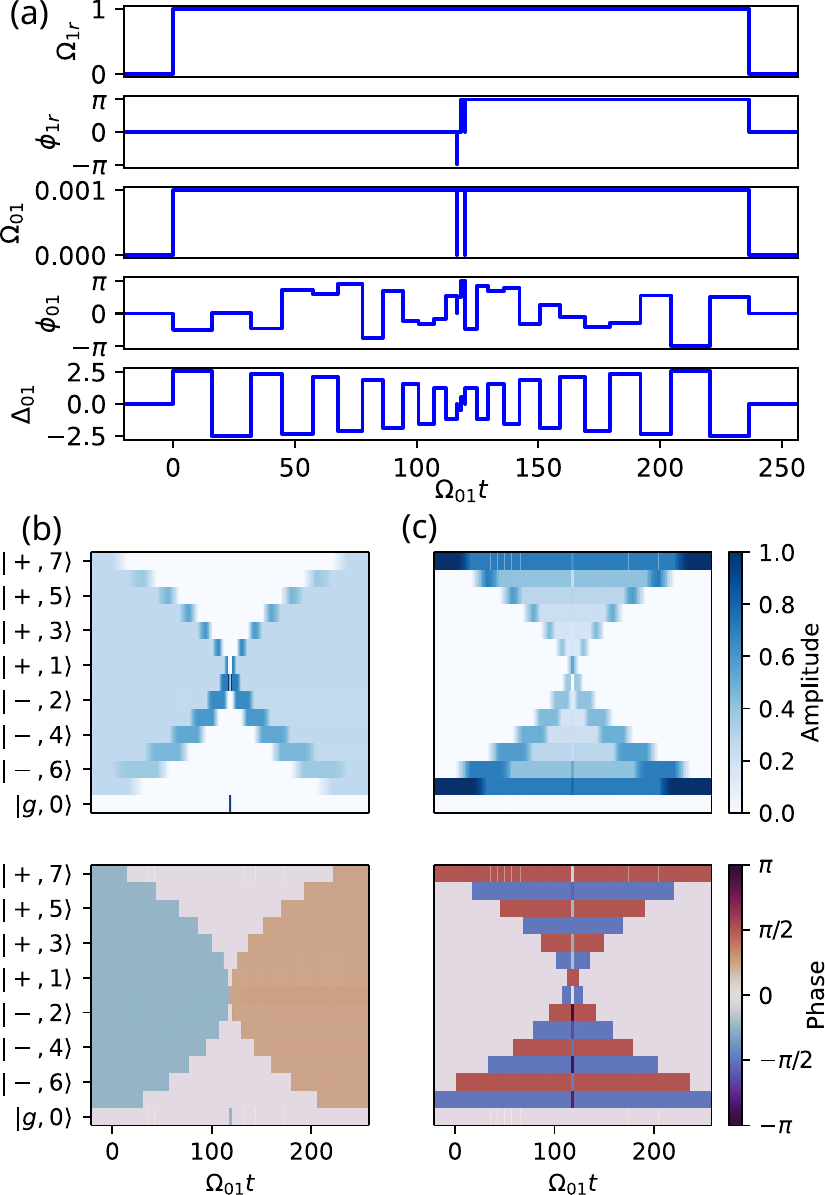}
	\caption{\textbf{Realization of a qudit phase gate.} \textbf{(a)} Pulse sequence for generating a phase gate with rotation angle $\pi/2$ on the target state $|\psi_{\text{target}}\rangle =  \frac{1}{\sqrt{2N}}\sum_\pm \sum_{q=1}^{N} |\pm,q\rangle$ for $N=7$ atoms ($14$-level qudit). \textbf{(b)} Time evolution of the initial state $|\psi_{\text{ini}}\rangle = e^{-i \pi/4} |\psi_{\text{target}}\rangle$ for the pulse sequence of (a) (we added a global phase of $-\pi/4$ to improve the visual rendering of the figure). The resulting final state is $|\psi_{\text{final}}\rangle = e^{i \pi/2} |\psi_{\text{in}}\rangle = e^{i \pi/4} |\psi_{\text{target}}\rangle$. We clearly notice the three phases during the dynamics: (i) the progressive mapping of the wave function to $|-,1\rangle$; (ii) in the middle of the pulse sequence, the $z$ rotation of angle $\pi/2$ acting on the subspace $\left\{|g,0\rangle, |-,1\rangle  \right\}$ that temporarily populates $|g,0\rangle$; and (iii) mapping back the wave function to $|\psi_{\text{final}}\rangle$. \textbf{(c)} The phase gate leaves the states orthogonal to $|\psi_{\text{target}}\rangle$ unchanged. To illustrate this point, we show the dynamics of $|\psi_{\text{ini}}\rangle = \frac{1}{\sqrt{2}} \left( e^{i \pi/2} |+,7\rangle + e^{-i \pi/2} |-,7\rangle  \right)$ that is unaffected by the pulse sequence, i.e., $|\psi_{\text{final}}\rangle = |\psi_{\text{ini}}\rangle$. Note that in (b) and (d) we subtracted the trivial phase accumulation due to the diagonal terms of the Hamiltonian (which is anyway compensated at the end of the pulse sequence by the protocol). To improve the readability of the figure, we have set the phases to zero when the corresponding amplitudes are below $10^{-3}$.}   \label{Fig2}
\end{figure}

\begin{figure}[t!]
\centering
	\includegraphics[width=\linewidth]{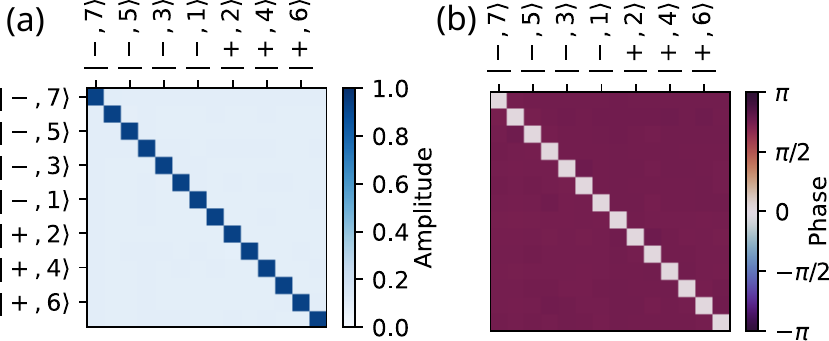}
	\caption{\textbf{Matrix representation of the generalized phase gate.} By simulating the time evolution of the pulse sequence of Fig. \ref{Fig2} (a) for each of the qudit basis vectors $\left\{ |\pm,q \rangle \right\}$ as initial state, we obtain the matrix representation of the generalized phase gate introduced in Fig. \ref{Fig2}. \textbf{(a)} Amplitude of the matrix. \textbf{(b)} Phase of the matrix. By comparing the gate $\hat U$ resulting from the simulation of the dynamics to the expected gate $\hat U_{\text{target}} = e^{i \pi/2} |\psi_{\text{target}}\rangle \langle\psi_{\text{target}}| + \left( \hat I - |\psi_{\text{target}}\rangle \langle\psi_{\text{target}}|\right)$, we extract a gate infidelity of $9 \times 10^{-5}$ for $\Omega_{01} / \Omega_{1r} = 10^{-3}$ for $N=7$ ($14$-level qudit) where the gate infidelity is defined as $\epsilon = 1 - \frac{1}{(2N)^2} \left| \text{Tr} \left( \hat U_{\text{target}}^\dagger \hat U \right) \right|^2$.} \label{Fig3}
\end{figure}

The general protocol to realize arbitrary unitaries is illustrated through two examples: the generalized phase gate and the generalized Hadamard gate.

\subsection{Generalized phase gate}

As a first example, we illustrate in Fig. \ref{Fig2} (a) a series of pulses to perform a $\pi/2$-phase gate on the target state $|\psi_{\text{target}}\rangle = \frac{1}{\sqrt{2N}}\sum_\pm \sum_{q=1}^{N} |\pm,q\rangle$, which we decompose as a series of three pulse sequences $\hat P_{|\psi_{\text{target}}\rangle, \pi/2} = \hat O^{-1}_{|\psi_{\text{target}}\rangle} \hat P_{|-,1\rangle, \pi/2}  \hat O_{|\psi_{\text{target}}\rangle}$ following the protocol of Sec. \ref{Gate_Protocol}. Throughout the entire paper, we simulate the ``exact'' time evolution of an initial state $|\psi_{\text{ini}}\rangle$ by matrix exponentiation
\begin{equation*}
	|\psi(t)\rangle = e^{-i \hat H^{(k)} \left(t - \sum_{p=1}^{k-1} T^{(p)}\right)} \prod_{p=1}^{k-1} e^{-i \hat H^{(p)} T^{(p)}} |\psi_{\text{ini}}\rangle
\end{equation*}
for $\sum_{p=1}^{k-1} T^{(p)} < t < \sum_{p=1}^{k} T^{(p)}$. The Hamiltonian $\hat H^{(k)}$ is the exact Hamiltonian $\hat H_0 + \hat H_{\text{c}}$ corresponding to the parameters of the $k$th pulse of the sequence. Note that since part of the pulse sequence is designed using an effective Hamiltonian $\hat H_{\text{eff}}$ that approximates $\hat H$, this protocol is not exact which results in gate errors discussed in Sec. \ref{Gate_Infidelities}.
Figure \ref{Fig2} (b) shows the dynamics of the system when the entire pulse sequence is applied to the initial state $|\psi_{\text{in}}\rangle = |\psi_{\text{target}}\rangle$. We can clearly identify (i) in the first part of the sequence, the role of $\hat O_{|\psi_{\text{target}}\rangle}$ that maps the initial wave function onto $|-,1\rangle$ while conserving the phase of the initial state; (ii) the effect of  $\hat P_{|-,1\rangle, \pi/2}$ that adds a phase $\pi/2$ to $|-,1\rangle$; and finally (iii) mapping back the wave function to the initial state  through the action of $\hat O^{-1}_{|\psi_{\text{target}}\rangle}$ but with an additional phase of $\pi/2$ that was acquired in (ii). The whole operation leads to $ \hat O^{-1}_{|\psi_{\text{target}}\rangle} \hat P_{|-,1\rangle, \pi/2}  \hat O_{|\psi_{\text{target}}\rangle} |\psi_{\text{target}}\rangle = e^{i\pi/2} |\psi_{\text{target}}\rangle$. Figure \ref{Fig2} (c) illustrates the effect of the pulse sequence applied to an initial state $|\psi_{\text{in}}\rangle = |\psi_\perp\rangle$ belonging to the subspace of $\mathcal H'$ orthogonal to $|\psi_{\text{target}}\rangle$. The action of $\hat O_{|\psi_{\text{target}}\rangle}$ leads to a state where $|-,1\rangle$ is not populated such that $\hat P_{|-,1\rangle, \pi/2}$ has no effect on this state. Finally, after the action of $\hat O^{-1}_{|\psi_{\text{target}}\rangle}$ we recover the initial wave function: $\hat O^{-1}_{|\psi_{\text{target}}\rangle} \hat P_{|-,1\rangle, \pi/2}  \hat O_{|\psi_{\text{target}}\rangle} |\psi_\perp\rangle = |\psi_\perp\rangle$. Finally, Fig. \ref{Fig3} gives the matrix representation of this gate in the dressed-state basis.

\subsection{Generalized Hadamard gate}

\begin{figure*}[t!]
\centering
\includegraphics[width=\linewidth]{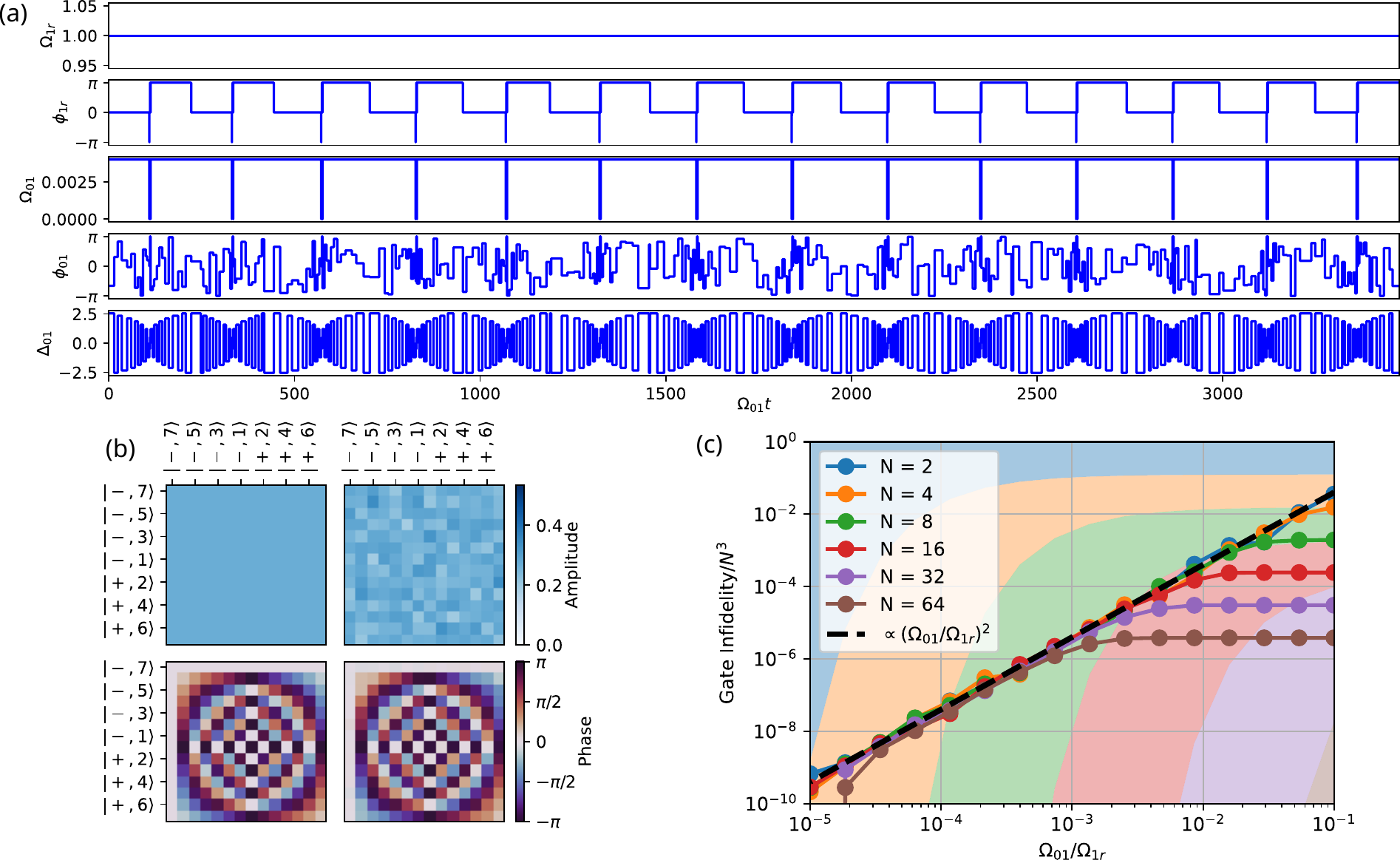}
	\caption{\textbf{Generalized Hadamard gate.} \textbf{(a)} Pulse sequence to generate the generalized Hadamard gate for $N=7$ atoms ($14$-level qudit) and $\Omega_{01}/\Omega_{1r} = 0.004$. It consists of repeatedly applying a phase gate with angle $\alpha_i$ to each of the $2N$ eigenvectors $|\psi_{\alpha_i}\rangle$ of the generalized Hadamard gate with eigenvalues $e^{i \alpha_i}$. \textbf{(b)} Left column: Amplitude and phase of the matrix representation of the target Hadamard gate for $N=7$ atoms. Right column: Matrix representation of the gate resulting from numerical time evolution of the pulse sequence given in (a) for each of the qudit basis vectors $\left\{ |\pm,q \rangle \right\}$ as initial state. For the sake of illustration, we have chosen $\Omega_{01}/\Omega_{1r} = 0.004$ so that the imperfections of the gate remain visible in the figure. For these parameters, the infidelity of the gate is $3 \times 10^{-2}$. \textbf{(c)} Numerical simulations of the infidelities of the generalized Hadamard gate as a function of $\Omega_{01}/\Omega_{1r}$ for different values of the number of atoms $N$ (the dimension of the qudit Hilbert space is $2N$). The results confirm the scaling of the infidelities $\sim N^3 \Omega_{01}^2 / \Omega_{1r}^2$. The shaded areas depict the regions where the probability of relaxation of the Rydberg state $P_{\text{relax}}$ exceeds the gate infidelity $\epsilon$ for different values of $N$ with the same color coding as the legend of the simulation data for $\Gamma_r^{-1} = 100 \, \mu \text{s}$ and $\Omega_{1r}/(2\pi) = 25 \, \text{MHz}$. It thus makes it possible to realize experimentally a generalized Hadamard gate with up to a $14$-level qudit ($N=7$) before being hindered by the lifetime of the Rydberg state.}   \label{Fig4}
\end{figure*}

We now turn to the case of a more complex gate by implementing the generalized Hadamard gate which is defined as
\begin{equation*}
	\hat U_{\text{Had}} : |q_j\rangle \mapsto \frac{1}{\sqrt{2N}} \sum_{p=1}^{2N} e^{i \frac{\pi}{N} (j-1)(p-1)} \, |q_p\rangle
\end{equation*}
where we have introduced the qudit basis such that $|q_1\rangle = |-,N\rangle, ..., |q_N\rangle = |-,1\rangle, |q_{N+1}\rangle = |+,1\rangle , ..., |q_{2N}\rangle = |+,N\rangle$. We start by diagonalizing $\hat U_{\text{Had}}$  
to find $2N$ eigenvalues and associated eigenvectors $\left\{ e^{i \alpha_i}, |\alpha_i\rangle\right\}$. The pulse sequence is then obtained by concatenating the pulse sequence corresponding to phase gates with angle $\alpha_i$ applied to the target vectors $|\alpha_i\rangle$ according to the decomposition $\hat U_{\text{Had}} = \prod_i \hat P_{|\alpha_i\rangle,\alpha_i}$. This pulse sequence is given in Fig. \ref{Fig4} (a) for $N = 7$ atoms (qudit with $14$ levels) where we can clearly spot the pattern of the phase gate sequence of Fig. \ref{Fig2} (a) repeated $2N$ times. Note that in the particular case where some of the $\alpha_i$'s are zero, we get $\hat P_{|\alpha_i\rangle,\alpha_i=0} = \hat I$ and the pulse sequence can be simplified: $\hat U_{\text{Had}} = \prod_{\alpha_i\neq 0} \hat P_{|\alpha_i\rangle,\alpha_i}$. However, we omitted this optimization in our code generating the pulse sequence of Fig. \ref{Fig4} to estimate the error rate of the gate in the most general case as discussed in the next section. The simulated matrix representation of the Hadamard gate is obtained by numerical integration of the dynamics associated to the pulse sequence applied to each of the dressed-state basis vectors as initial states. Figure \ref{Fig4} (b) compares the expected matrix representation of the Hadamard gate to the simulated one for a parameter regime of the simulation that keeps the imperfections of the gate visible to introduce the concept of gate infidelities. 

\section{Gate infidelities} \label{Gate_Infidelities}

The gate protocol to generate the pulse sequence is not exact as it relies on isolating a two-dimensional subspace in each pulse to perform a rotation with an effective Hamiltonian $\hat H_{\text{eff}}$. However, this procedure is only approximate as nonresonant terms of the Hamiltonian result in small corrections to the dynamics which decrease the gate fidelities. Moreover, the protocol relies on the upper laser to create the collective dressed states which we are used to encode the qudit, so intuitively the control laser should not disturb the initial energy-level structure of the qudit set by the laser addressing the upper transition (it should be a small perturbation) which means that we expect the protocol to work best in the regime $\Omega_{01}/\Omega_{1r} \ll 1$.

We define the gate infidelity as $\epsilon = 1 - \frac{1}{(2N)^2} \left| \text{Tr} \left( \hat U_{\text{target}}^\dagger \hat U \right) \right|^2$, where $\hat U_{\text{target}}$ is the ideal unitary gate we aim to create while $\hat U$ is the gate obtained from simulating the time evolution of the full Hamiltonian for the parameters coming from the pulse sequence. We interpret the gate infidelity as being the accumulation of errors occurring at each pulse of the sequence. Any nontrivial target phase gate requires $\sim N$ pulses while the Hadamard gate and more complex gates need the sequential application of $2N$ phase gates leading to a total number of pulses $\sim N^2$. The typical error accumulated during one pulse of the sequence involving an effective Hamiltonian is $\sim N \Omega_{01}^2 / \Omega_{1r}^2$ which leads to total infidelities of $\sim N^2 \Omega_{01}^2 / \Omega_{1r}^2$ for the generalized phase gate and $\sim N^3 \Omega_{01}^2 / \Omega_{1r}^2$ for the Hadamard gate. This simple scaling is in good agreement with numerical simulations shown in Fig \ref{Fig4} (c).

\section{Implementations and influence of the Rydberg state decay}

Natural candidates for implementing this protocol are atomic species with long-lived intermediate states. For example, alkali-metal atoms can be used to encode the $|0\rangle$ and $|1\rangle$ states in their hyperfine ground states with a microwave field or a pair of Raman beams to control the qudit. Using this approach, the first spectroscopic measurement of the dressed-state spectrum has already been reported in the experimental work Ref. \cite{Lee17}. Other promising candidates are alkaline-earth-like atoms such as strontium and ytterbium, for which the state $\ket{0}$ can be encoded in the ground state and $\ket{1}$ in a long-lived metastable state. This enables optical control via the clock transition, while the Rydberg state $\ket{r}$ can be accessed from $\ket{1}$ through single-photon excitation using a UV laser \cite{Tsai25}. An alternative promising approach is to use the nuclear spin of the long-lived metastable state to encode the states $\ket{0}$ and $\ket{1}$ \cite{Ma23}. The complex pulse sequences of the protocol can be realized with arbitrary waveform generators in the microwave domain or using  arbitrary optical pulse shapers in the optical domain \cite{Evered23,Ma23,Yang25}.

One issue which might hinder the applicability of the presented qudit control protocols is the decay of the Rydberg state with a lifetime of $\approx 100 \, \mu\text{s}$  which limits the maximum duration of the pulse sequence. We can estimate the decay probability of a state $|\psi(t)\rangle$ over a given pulse sequence
\begin{equation*}
        P_{\text{relax}} = \exp \left[- \Gamma_r \int_{\text{pulse seq.}} \mathrm d t \, \langle \psi(t)|\hat P_r|\psi(t)\rangle \right]
\end{equation*}
where $\hat P_r = \sum_j |r_j\rangle\langle r_j|$ is the projector on the Rydberg state manifold and $\Gamma_r$ is the decay rate of the Rydberg state. Since along a pulse sequence the state $|\psi(t)\rangle$ explores \emph{on average} as many states with and without Rydberg excitation, it is safe to assume that $\int_{\text{pulse seq.}} \mathrm d t \, \langle \psi(t)|\hat P_r|\psi(t)\rangle \simeq 0.5 \, T_{\text{tot}}$ where $T_{\text{tot}}$ is the total duration of the pulse sequence. The scaling of the total pulse sequence duration for arbitrary state preparation is, in the most pessimistic scenario, $T_{\text{tot}} \sim N / \Omega_{01}$ while for a generalized phase gate $T_{\text{tot}} \sim N^2 / \Omega_{01}$ and for a generalized Hadamard gate (or arbitrary complex gates) $T_{\text{tot}} \sim N^3 / \Omega_{01}$. Focusing first on the case of the generalized Hadamard gate, we find that $P_{\text{relax}} \simeq \exp \left[- 0.5 \, \Gamma_r  N^3 \frac{1}{\Omega_{1r}} \frac{\Omega_{1r}}{\Omega_{01}}   \right]$. 

As expected, larger values of $\Omega_{1r}$ enable faster operations (for a given fidelity, which depends on the ratio $\Omega_{01}/\Omega_{1r}$) and thus minimize the probability of Rydberg decay. Experimentally, large values of $\Omega_{1r}$ can be achieved with the availability of high-power lasers, using either single-photon \cite{Ma23, Scholl23, Muniz25} or two-photon excitation schemes \cite{Evered23}. However, the value of $\Omega_{1r}$ must be maintained well below the Rydberg interaction strength in order to satisfy the hard-blockade condition. For realistic Rydberg interaction strengths, we set $\Omega_{1r}/(2\pi) = 25\,\text{MHz}$ and use a typical lifetime $\Gamma_r^{-1} = 100\,\mu\text{s}$ to plot in Fig.~\ref{Fig4}(c) the regions where the relaxation probability $P_{\text{relax}}$ is smaller than the gate infidelity $\epsilon$. From this plot, we conclude that it should be possible to realize a generalized Hadamard gate (or any complex gate) with up to a 14-level qudit ($N=7$ atoms). For qudits with more levels (more atoms), the decay of the Rydberg state will impede the realization of complex unitaries. Generalized phase gates, having a much shorter pulse sequence, can be realized with up to a 24-level qudit (12 atoms) while we estimate that arbitrary state preparation can be realized with up to a 340-level qudit (170 atoms) before the decay of the Rydberg state precludes the possibility to realize these operations.   

In order to overcome these limitations, one can think of embedding the system in a cryogenic environment at a temperature of a few Kelvin which would enhance the lifetime of low angular momentum Rydberg states by a factor $2$ to $3$ \cite{Beterov09}. This would help get better gate or state preparation fidelities but would not be a real game changer in terms of the size of the qudit Hilbert space that can be realistically manipulated. 

In addition to the spontaneous decay of the Rydberg state, experimental implementations are likely to face further challenges. As previously noted, errors arising from fluctuations in the collective parameters of the multilevel, e.g., $K_N^q$ and $Q_N^q$, can be mitigated by using a well-controlled number of physical atoms to compose the Rydberg superatom. This can be achieved using mature approaches such as arrays of single atoms in optical tweezers, rather than relying on atomic ensembles~\cite{Mei22} or Mott insulators~\cite{Zeiher15}, where the number of atoms involved may fluctuate (even though Mott insulators exhibit sub-Poissonian atom number fluctuations).

The impact of additional experimental noise sources -- such as laser phase noise or inhomogeneities, spontaneous emission from intermediate states (particularly in two-photon Rydberg excitation schemes), and atomic motion (e.g., Doppler shifts) -- has been systematically studied in the context of single-atom manipulation~\cite{deLeseleuc18} and few-body assemblies~\cite{Lee19}. Extending these studies to the present protocol, and validating them experimentally, would provide critical insights into the feasibility of large-scale implementation.

\section{Conclusion and outlook}

We presented a protocol to use a Rydberg-blockaded array of three-level atoms as a qudit compatible with arbitrary state synthesis and unitary operations. This is achieved through engineering a pulse sequence of a control laser to manipulate the Jaynes-Cummings-like dressed states of a Rydberg-blockaded array of three-level atoms which emerge from a laser coupling the intermediate to the Rydberg states. We found that the best state preparation and gate fidelities result from a tradeoff on the total time of the sequence between a small ratio of the parameter $\left(\Omega_{01}/\Omega_{1r}\right)^2$ which favors longer pulse sequences and the lifetime of the Rydberg state whose effect is minimized by shortening the pulse sequence.

Possible extensions of this work include the development of entangling two qudit gates \cite{Gonzalez22,Omanakuttan23} which would enable universal quantum computation with this platform by spatially arranging Rydberg-blockaded arrays to serve as a large qudit register. In addition, exploring strategies to parallelize the operations for manipulating the qudit \cite{OLeary06} could significantly boost the speed of the whole procedure and enable a more flexible manipulation of larger qudits. Another promising direction is the integration of optimal control techniques to design pulse sequences, which can minimize sequence duration, smooth temporal variations of the control parameters, and mitigate experimental imperfections \cite{Jandura22,Pagano22,Evered23,Ma23}, thereby enhancing the robustness and scalability of the protocol in future implementations. 

We also envision exploiting the qudit Hilbert space for quantum computing permutation invariant codes where the two logical qubits, which are linear superpositions of Dicke states, belong to the qudit Hilbert space. This kind of code can correct for deletion errors (unlike other codes) and they have transversal non-Clifford gates \cite{Aydin24,Kubischta25}. Our protocol can also be applied to another type of permutation invariant code called a GNU code for error corrected quantum sensing \cite{Ouyang22}. Our protocol enables the preparation and manipulation of arbitrary symmetric Dicke states \cite{Lukin01} (superpositions of dressed states), which have applications in quantum metrology and the exploration of entanglement in macroscopic quantum systems. Finally, this platform can shine light on quantum simulation of Jaynes-Cummings and Jaynes-Cummings-Hubbard Hamiltonians when several of these systems are coupled together, hence realizing atomic analogs of these models.

\begin{acknowledgments}
We are grateful to Shannon Whitlock for insightful discussions and helpful comments on the manuscript. We thank Gavin Brennen for stimulating discussions and suggestions.
This work of the Interdisciplinary Thematic Institute QMat, as part of the ITI 2021-2028 program of the University of Strasbourg, Centre National de la Recherche Scientifique and Inserm, was supported by IdEx Unistra (Grant No. ANR-10-IDEX-0002) and by the SFRI STRAT’US project (Grant No. ANR-20-SFRI-0012) and EUR QMAT Grant No. ANR-17-EURE-0024 under the framework of the French Investments for the Future Program. In addition, this work has benefited from the financial support of Agence Nationale de la Recherche, France, under Grant No. ANR-22-CE30-0043 SIQ-FLight.
\end{acknowledgments}

\appendix

\section{Derivation of the Hamiltonian in the dressed-state basis} \label{Appendix_A}

In this appendix, we derive the Hamiltonian in the dressed-state basis for the Rydberg-blockaded array of atoms described in the main text. Throughout this appendix, we restrict ourselves to the fully symmetric subspace, which is dynamically invariant under the driving terms considered here. We begin by introducing the collective operators and symmetric states, and then express the bare and control Hamiltonians in the dressed-state basis.

We first introduce the individual spin operators
\begin{align*}
    \hat J^{1r}_{j+} &= |r_j\rangle\langle 1_j|, & \hat J^{1r}_{j-} &= |1_j\rangle\langle r_j|, \\
    \hat J^{01}_{j+} &= |1_j\rangle\langle 0_j|, & \hat J^{01}_{j-} &= |0_j\rangle\langle 1_j|, \\
	\hat J^{0r}_{j+} &= |r_j\rangle\langle 0_j|, & \hat J^{0r}_{j-} &= |0_j\rangle\langle r_j|,
\end{align*}
and the corresponding collective spin operators
\begin{align*}
    \hat J_{N+}^{1r} = \sum_j \hat J_{j+}^{1r}, \quad
	\hat J_{N-}^{1r} = \sum_j \hat J_{j-}^{1r}, \\
	 \hat J_{N+}^{01} = \sum_j \hat J_{j+}^{01}, \quad
        \hat J_{N-}^{01} = \sum_j \hat J_{j-}^{01}, \\
	\hat J_{N+}^{0r} = \sum_j \hat J_{j+}^{0r}, \quad
        \hat J_{N-}^{0r} = \sum_j \hat J_{j-}^{0r} \, .
\end{align*}

We also define the individual projectors
\begin{align*}
    \hat P^{r}_{j} &= |r_j\rangle\langle r_j|, & \hat P^{1}_{j} &= |1_j\rangle\langle 1_j|,
\end{align*}
and the corresponding collective projectors
\begin{align*}
    \hat P^{r}_{N} &= \sum_j \hat P^{r}_{j}, & \hat P^{1}_{N} &= \sum_j \hat P^{1}_{j}.
\end{align*}

In addition, we define the collective symmetric states without Rydberg excitation
\begin{align*}
    |g,q\rangle &\equiv C_q^N \, \bigl( \hat J_{N+}^{01} \bigr)^q \, |0\rangle^{\otimes N} \\
    &\equiv \{\ket{0}^{\otimes N-q}\ket{1}^{\otimes q}\}_{\text{sym}}, 
    \quad q = 0,\dots,N,
\end{align*}
where the normalization factor is
\begin{equation*}
    C_q^N = \sqrt{\frac{(N-q)!}{N!\,q!}}.
\end{equation*}
We further introduce the collective states with one Rydberg excitation
\begin{align*}
    |e,q\rangle 
    &\equiv \frac{1}{\sqrt{N-q}} \, \hat J_{N+}^{0r} \, |g,q\rangle \\
    &\equiv \{ \ket{0}^{\otimes (N-q-1)} \ket{1}^{\otimes q} \ket{r} \}_{\mathrm{sym}},
    \qquad q = 0,\dots,N-1 ,
\end{align*}
which can equivalently be written as
\begin{equation*}
    |e,q\rangle = \frac{1}{\sqrt{q+1}} \, \hat J_{N+}^{1r} \, |g,q+1\rangle .
\end{equation*}
The dressed states are defined as symmetric and antisymmetric superpositions of two collective states with $q$ excitations:
\begin{equation}
	|\pm,q\rangle = \frac{1}{\sqrt 2} \left( |e,q-1\rangle \pm |g,q\rangle \right) \, . \label{Dressed_States_Expression} 
\end{equation}

The action of the collective spin operators on the symmetric states reads
\begin{align*}
    \hat J_{N+}^{01} \, |g,q\rangle &= \sqrt{(N-q)(q+1)} \, |g,q+1\rangle, \\
    \hat J_{N-}^{01} \, |g,q\rangle &= \sqrt{(N-q+1)q} \, |g,q-1\rangle, \\
    \hat J_{N+}^{01} \, |e,q\rangle &= \sqrt{(N-q-1)(q+1)} \, |e,q+1\rangle, \\
    \hat J_{N-}^{01} \, |e,q\rangle &= \sqrt{(N-q)q} \, |e,q-1\rangle .
\end{align*}
Similarly, the operators $\hat J_{N\pm}^{1r}$ act as
\begin{align*}
    \hat J_{N+}^{1r} \, |g,q\rangle &= \sqrt{q} \, |e,q-1\rangle, \\
    \hat J_{N-}^{1r} \, |g,q\rangle &= 0, \\
    \hat J_{N+}^{1r} \, |e,q\rangle &= 0, \\
    \hat J_{N-}^{1r} \, |e,q\rangle &= \sqrt{q+1} \, |g,q+1\rangle .
\end{align*}

Using these notations, the bare system Hamiltonian, Eq.~(\ref{H_0}), can be written as
\begin{align*}
    \hat H_0 &= \frac{\Omega_{1r}}{2} \left( e^{-i \phi_{1r}} \hat J_{N+}^{1r} +  e^{i \phi_{1r}}  \hat J_{N-}^{1r} \right) \\
    &\quad + \frac{1}{2} \sum_j \sum_{k \neq j} V_{jk} \hat P_j^r \hat P_k^r \,,
\end{align*}
and the control Hamiltonian, Eq.~(\ref{H_c}), takes the form
\begin{align*}
    \hat H_{\text{c}} &= \frac{\Omega_{01}}{2} \left( e^{-i \phi_{01}} \hat J_{N+}^{01} + e^{i \phi_{01}} \hat J_{N-}^{01} \right) \notag \\
    &\quad - \Delta_{01} \left( \hat P_{N}^{1} + \hat P_{N}^{r} \right) \,.
\end{align*}

In the hard-blockade limit, the Hamiltonian is restricted to the sector with at most one Rydberg excitation,
\begin{align} \label{H_0_HB}
    \hat H_0^{\text{HB}}
    &= \hat \Pi \hat H_0 \hat \Pi \notag \\
    &= \frac{\Omega_{1r}}{2} \, \hat \Pi
    \left( e^{-i \phi_{1r}} \hat J_{N+}^{1r} + e^{i \phi_{1r}} \hat J_{N-}^{1r} \right)
    \hat \Pi ,
\end{align}
where $\hat \Pi$ denotes the projector onto the subspace with at most one atom in the Rydberg state $\ket{r}$. In this limit, the interaction term proportional to $V_{jk}$ vanishes identically since states with more than one Rydberg excitation are projected out.
Similarly, the control Hamiltonian reduces to
\begin{equation} \label{Hc_appendix}
    \hat H_{\text{c}}^{\text{HB}} = \hat \Pi \hat H_{\text{c}} \hat \Pi .
\end{equation}

In the following, we always work in the hard-blockade limit and therefore omit the superscript ``HB'' in the following for notational simplicity. 
Within this approximation, the Hamiltonian can be written equivalently in several bases: the bare-state basis
$\{\bigotimes_j |\alpha_j\rangle\}$ with $|\alpha_j\rangle \in \{|0_j\rangle,|1_j\rangle,|r_j\rangle\}$,
the symmetric basis
$\{|g,0\rangle,\ldots,|g,N\rangle, |e,0\rangle,\ldots,|e,N-1\rangle\}$,
or the dressed-state basis
$\{|g,0\rangle,|\pm,q\rangle\}_{q=1,\ldots,N}$.
Note that the latter two bases have the same dimension, $2N+1$.

We now express the bare Hamiltonian $\hat H_0$ in the symmetric and dressed-state bases. Substituting the expressions of the collective spin operators
\begin{align*}
	\hat J_{N+}^{1r} &= \sum_{q=1}^N \sqrt q |e,q-1\rangle \langle g,q|, \\
	\hat J_{N-}^{1r} &= \sum_{q=1}^N \sqrt q |g,q\rangle \langle e,q-1|
\end{align*}
into Eq. (\ref{H_0_HB}), we get the bare system Hamiltonian in the symmetric state basis:
\begin{align*}
	\hat H_0 = & \frac{\Omega_{1r}}{2} \, \sum_{q=1}^{N} \sqrt{q}  \Bigl( e^{-i \phi_{1r}}  |e,q-1\rangle\langle g,q|  \notag \\
	&\qquad  \qquad  \qquad  + e^{i \phi_{1r}} |g,q\rangle\langle e,q-1| \Bigr). 
\end{align*}
Using the definition (\ref{Dressed_States_Expression}), $\hat H_0$ can be rewritten in the dressed-state basis as
\begin{align*}
\hat H_0 &= \sum_{q=1}^{N} \frac{\Omega_{1r} \sqrt{q}}{2} \Bigl[
    \cos\phi_{1r} \bigl( |+,q\rangle \langle +,q| - |-,q\rangle \langle -,q| \bigr) \notag \\
&\qquad + i \sin\phi_{1r} \bigl( |+,q\rangle \langle -,q| - |-,q\rangle \langle +,q| \bigr)
\Bigr] \\
&= \sum_{q=1}^{N} \frac{\Omega_{1r} \sqrt{q}}{2} \, 
\mathbf n_{\phi_{1r}} \cdot \widehat{\boldsymbol{\sigma}}_{\{|+,q\rangle,|-,q\rangle\}} \, ,
\end{align*}
where the vectorial Pauli operators $\widehat{\boldsymbol{\sigma}}_{\left\{|u\rangle,|v\rangle \right\}} = \left[\hat \sigma_x, \hat \sigma_y, \hat \sigma_z\right]$ act on the subspace $\left\{|u\rangle,|v\rangle \right\}$, and the unit vector is $\mathbf n_{\phi_{1r}} = (0, - \sin \phi_{1r},\cos \phi_{1r})$. We thus recover the expression of Eq. (\ref{H_0_dressed_states}) of the main text.

We now repeat the same procedure for the control Hamiltonian $\hat H_{\text{c}}$. In particular, in the symmetric-state basis, the collective raising operator can be written as
\begin{align*}
\hat J_{N+}^{01} &= \sum_{q=0}^{N-2} \sqrt{(N-q-1)(q+2)} \, |g,q+2\rangle \langle g,q+1| \\
&\quad + \sum_{q=0}^{N-2} \sqrt{(N-q-1)(q+1)} \, |e,q+1\rangle \langle e,q| \\
&\quad + \sqrt{N} \, |g,1\rangle \langle g,0| \,.
\end{align*}

Using the definition (\ref{Dressed_States_Expression}), we can now express $\hat J_{N+}^{01}$ in the dressed-state basis:
\begin{align} \label{J01}
\hat J_{N+}^{01} &= \sum_{q=1}^{N-1} \frac{\sqrt{N-q}}{2} \Bigl[
    (\sqrt{q+1} + \sqrt{q}) \, |+,q+1\rangle \langle +,q| \notag\\
&\quad + (\sqrt{q+1} + \sqrt{q}) \, |-,q+1\rangle \langle -,q| \notag\\
&\quad - (\sqrt{q+1} - \sqrt{q}) \, |+,q+1\rangle \langle -,q| \notag\\
&\quad - (\sqrt{q+1} - \sqrt{q}) \, |-,q+1\rangle \langle +,q|
\Bigr] \notag\\
&\quad + \sqrt{\frac{N}{2}} \left( |+1\rangle \langle g,0| - |-1\rangle \langle g,0| \right) .
\end{align}

Similarly, the operator $\hat J_{N-}^{01}$ can be obtained in the symmetric or dressed-state basis by taking the Hermitian conjugate of $\hat J_{N+}^{01}$. We omit its explicit expression to avoid overloading the text.

Moreover, we also get in the symmetric and dressed-state bases the following relations:
\begin{align} \label{Projector}
	\hat P_{N}^{1} + \hat P_{N}^{r} & = \sum_{q=1}^{N} \, q \, \left(  |g, q\rangle \langle g, q| +  |e, q-1\rangle \langle e, q-1|  \right) \notag \\ 
	& = \sum_{\pm} \sum_{q=1}^{N} \, q \,  |\pm, q\rangle \langle \pm, q| \, . 
\end{align}
Substituting the expressions Eqs. (\ref{J01}) and (\ref{Projector}) into Eq. (\ref{Hc_appendix}), we can obtain the expression of the control Hamiltonian in the dressed-state basis:
\begin{align*} 
\hat{H}_{\text{c}} ={} & \frac{\Omega_{01}}{2} \Biggl[ \sum_{\pm} \sum_{q=1}^{N-1} K_{N}^{q} \, \mathbf{n}_{\phi_{01}} \cdot \widehat{\boldsymbol{\sigma}}_{\{|\pm,q+1\rangle,|\pm,q\rangle\}} \notag \\
& \quad - \sum_{\pm} \sum_{q=1}^{N-1} Q_{N}^{q} \, \mathbf{n}_{\phi_{01}} \cdot \widehat{\boldsymbol{\sigma}}_{\{|\pm,q+1\rangle,|\mp,q\rangle\}} \notag \\
& \quad + \sqrt{\frac{N}{2}} \, \mathbf{n}_{\phi_{01}} \cdot \widehat{\boldsymbol{\sigma}}_{\{|+1\rangle,|g,0\rangle\}} \notag \\
	& \quad - \sqrt{\frac{N}{2}} \, \mathbf{n}_{\phi_{01}} \cdot \widehat{\boldsymbol{\sigma}}_{\{|-1\rangle,|g,0\rangle\}} \Biggr] \notag \\
& - \Delta_{01} \sum_{\pm} \sum_{q=1}^N q \, |\pm,q\rangle \langle \pm,q| \, ,
\end{align*}
where $K_N^q = \frac{\sqrt{(N-q)}}{2 \left(\sqrt{q+1} - \sqrt{q}\right) }$, $Q_N^q = \frac{\sqrt{(N-q)}}{2 \left(\sqrt{q+1} + \sqrt{q}\right)}$, and $\mathbf n_{\phi_{01}} = (\cos \phi_{01}, \sin \phi_{01},0)$. We thus recover the expression of Eq. (\ref{H_c_dressed_states}) of the main text. An explicit expression of $\hat H_c$ in the symmetric basis can be obtained similarly by substituting Eqs.~(\ref{J01}) and (\ref{Projector}) into Eq.~(\ref{Hc_appendix}).

\section{Derivation of the Effective Hamiltonians} \label{Appendix_B}

In this appendix, we show how to derive the effective Hamiltonians presented in Eqs.~(\ref{eq:Heff}) and (\ref{Heff_Phase_Gate}) of the main text. We provide a detailed derivation for Eq.~(\ref{eq:Heff}); the expression for Eq.~(\ref{Heff_Phase_Gate}) is obtained using the same method.

\subsection{Effective Hamiltonian for $\phi_{1r}=0$}

Setting $\phi_{1r} = 0$ leads to a diagonal bare system Hamiltonian, Eq.~(\ref{H_0_dressed_states}), when expressed in the dressed-state basis:
\begin{equation*}
	\hat H_0 = \sum_\pm \sum_{q=1}^N E_{\pm,q} \ket{\pm,q}\bra{\pm,q} \, ,
\end{equation*}
where
\begin{equation*}
	E_{\pm,q} = \pm \frac{\Omega_{1r} \sqrt{q}}{2} \quad \text{for} \quad \phi_{1r} = 0 \, .
\end{equation*}
In this form, the bare system Hamiltonian is analogous to the Jaynes-Cummings Hamiltonian~\cite{Jaynes63, Keating16}. The total Hamiltonian, Eq.~(\ref{Htot}), then takes the following form in the dressed-state basis:
\begin{align*}
\hat{H} ={} & \frac{\Omega_{01}}{2} \Biggl[ \sum_{\pm} \sum_{q=1}^{N-1} K_{N}^{q} \, \mathbf{n}_{\phi_{01}} \cdot \widehat{\boldsymbol{\sigma}}_{\{|\pm,q+1\rangle,|\pm,q\rangle\}} \notag \\
& \quad - \sum_{\pm} \sum_{q=1}^{N-1} Q_{N}^{q} \, \mathbf{n}_{\phi_{01}} \cdot \widehat{\boldsymbol{\sigma}}_{\{|\pm,q+1\rangle,|\mp,q\rangle\}} \notag \\
& \quad + \sqrt{\frac{N}{2}} \, \mathbf{n}_{\phi_{01}} \cdot \widehat{\boldsymbol{\sigma}}_{\{|+1\rangle,|g,0\rangle\}} \notag \\
	&  \quad - \sqrt{\frac{N}{2}} \, \mathbf{n}_{\phi_{01}} \cdot \widehat{\boldsymbol{\sigma}}_{\{|-1\rangle,|g,0\rangle\}} \Biggr] \notag \\
        & + \sum_{\pm} \sum_{q=1}^N  \left[ \pm \frac{\Omega_{1r} \sqrt{q}}{2}  - \Delta_{01} q   \right] \, |\pm,q\rangle \langle \pm,q| \,.
\end{align*}
In this representation, the last line of the Hamiltonian is diagonal in the dressed-state basis, while the terms above are off-diagonal.

To isolate the subspace $\left\{|\pm,q+1\rangle,|\mp,q\rangle \right\}$, we need to adjust the detuning of the control laser to match the energy difference between these two states: 
\begin{align*}
        \Delta_{01} & = E_{\pm,q+1} - E_{\mp,q} \\
        & =  \pm \frac{\Omega_{1r}}{2\left( \sqrt{q+1} - \sqrt{q} \right) } \, . 
\end{align*}
When this detuning is set, and assuming $\Omega_{01} \ll \Omega_{1r}$, we neglect all the off-diagonal elements of the Hamiltonian outside of the $\left\{|\pm,q+1\rangle,|\mp,q\rangle \right\}$ subspace to get
\begin{align*}
    \hat{H}_{\text{eff}} ={} & - \frac{\Omega_{01} Q_N^q}{2} \, \mathbf{n}_{\phi_{01}} \cdot \widehat{\boldsymbol{\sigma}}_{\{|\pm,q+1\rangle,|\mp,q\rangle\}} \notag \\
    & + \sum_\pm \sum_{q=1}^N \left( \pm \frac{\Omega_{1r} \sqrt{q}}{2}  - \Delta_{01} q \right) \, |\pm,q\rangle \langle \pm,q| 
\end{align*}
which is the expression of the effective Hamiltonian Eq.~(\ref{eq:Heff}) given in the main text.

\subsection{Effective Hamiltonian for $\phi_{1r}=\pi$}

An analogous reasoning applies when $\phi_{1r} = \pi$. In that case, 
\begin{equation*}
        E_{\pm,q} = \mp \frac{\Omega_{1r} \sqrt{q}}{2} \quad \text{for} \quad \phi_{1r} = \pi \, , 
\end{equation*}
and we need to fix the detuning to
\begin{align*}
        \Delta_{01} & = E_{\pm,q+1} - E_{\mp,q} \\
        & =  \mp \frac{\Omega_{1r}}{2\left( \sqrt{q+1} - \sqrt{q} \right) } 
\end{align*}
to obtain the effective Hamiltonian
\begin{align*}
    \hat{H}_{\text{eff}} ={} & - \frac{\Omega_{01} Q_N^q}{2} \, \mathbf{n}_{\phi_{01}} \cdot \widehat{\boldsymbol{\sigma}}_{\{|\pm,q+1\rangle,|\mp,q\rangle\}} \notag \\
    & + \sum_\pm \sum_{q=1}^N \left( \mp \frac{\Omega_{1r} \sqrt{q}}{2}  - \Delta_{01} q \right) \, |\pm,q\rangle \langle \pm,q| \, . 
\end{align*}

\subsection{Phase cancellation protocol}

By changing $\phi_{1r}$ from zero to $\pi$, the sign of the diagonal elements of $\hat{H}_{\text{eff}}$ is inverted. This feature can be exploited during qudit manipulation in the $\{|\pm,q+1\rangle,|\mp,q\rangle\}$ subspace to cancel the phase accumulated by dressed states outside the manipulation subspace. For example, this can be achieved by performing half of the rotation with $\phi_{1r}=0$ and the other half with $\phi_{1r}=\pi$, as explained in the main text.

An alternative approach, also discussed in the main text, consists of performing the entire first part of the pulse sequence—corresponding to full control over the Hilbert space—with $\phi_{1r}=0$, then applying the phase gate, and finally canceling the accumulated phase by setting $\phi_{1r}=\pi$ during the last stage, when the full control over the Hilbert space is reversed. This is the approach adopted in our numerical simulations.

\end{document}